\documentstyle[12pt]{report}
\let\ssection=\section
\renewcommand{\section}{\setcounter{equation}{0}\ssection}
\newcommand{\BE}{\begin{equation}}
\newcommand{\EE}{\end{equation}}
\newcommand{\BA}{\begin{eqnarray}}
\newcommand{\EA}{\end{eqnarray}}

\def\a{\alpha}
\def\b{\beta}

\def\ltextindent#1{\hbox to \hangindent{#1\hss}\ignorespaces}
\def\sqr#1#2{{\vcenter{\hrule height.#2pt\hbox{\vrule width.#2pt
height#1pt \kern#1pt \vrule width.#2pt}\hrule height.#2pt}}}

\def\negenspace{\kern-1.1em}

\def\negenspaceexp{\kern-0.5em}

\begin{document}
\section{Introduction: the need to go beyond Riemannian manifolds}

A number of developments in physics in recent years have evoked the
possibility that the treatment of spacetime might involve more than
just the Riemannian spacetime of Einstein's general relativity:
\begin{description}

\item[(1)] The vain effort so far to {\it quantize gravity} is, perhaps,
the strongest piece of evidence for going beyond a
geometry which is dominated by the classical distance concept.

\item[(2)] Intuitively, the  generalization of the three--dimensional
theory of elastic {\it continua with microstructure} to the
four--dimensional spacetime of gravity suggests, in a rather
convincing manner, physical interpretations for the newly emerging
structures in post--Riemannian spacetime geometry.

\item[(3)] The description of hadron (or nuclear) matter in terms of
{\it extended structures}. In particular, the quadrupole pulsation rates
of that matter and, in a rest frame, their relation to representations of
the volume--preserving
three--dimensional linear group $SL(3,R)$ --- with the rotation group
$SO(3)$ as subgroup  --- have been established experimentally.

\item[(4)] The study of the {\it early universe} --- in the light of
the various theorems about a singular origin, the ideas about
unification of the fundamental interactions (mostly involving
additional dimensions, later compactified) and inflationary models
with dilaton--induced Weyl covector ---
\end{description}

\noindent...and each of these developments necessitates the
study of dynamical theories involving post--Riemannian geometries,
whether in the context of local field theories or within the framework
of string theories.  We explain the interest in continua with
microstructure, in extended structures, and the problematics of the
early universe -- as far as these are relevant as motivations for a
relaxation of the Riemannian constraint in gravity -- in the rest of
this chapter, leaving the rather involved issue of quantum gravity to
Chap.2.

The smallest departure from a $V_{4}$ would consist in admitting {\it
torsion}, the field strength of local translations, arriving thereby
at a Riemann-Cartan spacetime $U_{4}$ and, furthermore, {\it
nonmetricity}, resulting in a {\it metric-affine} $(L_{4},g)$
spacetime [He6]. In what follows, starting with Chapter 3, we deal
with the geometry of spacetime, with the Euler-Lagrange field
equations, with the Noether conservation identities, with the
conformal properties, and with a specific model of spontaneous
symmetry breakdown, in the framework of such metric-affine spacetimes.
For reasons that will become clear in the sequel, we study in
particular spacetime models arising from a Weyl/Yang-Mills-like gauge
theory approach to gravity.

\section{Spacetime as a continuum with microstructure}

In Einstein's general relativity theory (GR), the linear connection of
its Riemannian spacetime is (i) metric(--compatible), that is, the
length and angle measurements are integrable, and (ii) symmetric. The
symmetry of the Riemannian (or Levi-Civita) connection translates into
the closure of infinitesimally small parallelograms. Already the
transition from the flat gravity--free Minkowski spacetime to the
Riemannian spacetime in Einstein's theory can locally be understood as
a {\it deformation} process. A strain tensor $\varepsilon_{AB}$ in
continuum mechanics [La3] measures by its very definition
$\varepsilon_{AB}:=(\stackrel{\rm defor.}{g_{AB}} -\stackrel{\rm
undef.}{g_{AB}})/2$ the change of the metric between the undeformed
and the deformed state. Thus, because of the pairing of stress and
strain, it does not come as a surprize that in GR, according to
Hilbert's definition, the {\it stress}-energy-momentum tensor couples
the Lagrangian to the {\it metric}.

The lifting of the constraints of metric--compatibility and symmetry
yields nonmetricity and torsion, respectively. The continuum under
consideration, here classical spacetime, is thereby assumed to have a
non-trivial microstructure, similar to that of a liquid crystal or a
dislocated metal or the like. In particular, to drop the metricity
condition, i.e.\ to allow for nonmetricity $Q_{\a\b}:=-Dg_{\a\b}\neq
0$, and to ``touch'' thereby the lightcone, if parallelly displaced,
is classically a step of unusual boldness, but may be unavoidable in
quantum gravity. It is gratifying, though, to have the geometrical
concepts of nonmetricity and torsion already arising in the
(three--dimensional) continuum theory of lattice defects -- and there
they have concrete interpretations as densities of point defects and
line defects (dislocations), respectively, cf.\ [Kr2, Kr4].  But even
more, certain types of ``hyperstress'' are induced by these
post--Riemannian structures: Double--stress without moment relates to
nonmetricity, spin moment stress to torsion.\footnote{There exists an
extended literature on continua with microstructure, see, for example
Jaunzemis [Ja2], Mindlin [Mi15], and Nye [Ny1]. Kr\"oner's articles
[Kr1, Kr2, Kr3, Kr4, Kr5, He12] on lattice defects are particularly
illuminating, since they relate differential geometric notions to
distributions of lattice defects. His article on the lattice
interpretation of nonmetricity [Kr5] seems remarkable; however, no use
of it has been made so far.  The gauge--theoretical point of view is
stressed by Kleinert [Kl1]. The analogies between three-- and
four--dimensional continua with microstructure have been particularly
worked out by [Gr8, He1, He22, Mc2].}

Just as ordinary stress is the analogue of the (Hilbert)
energy--momentum density, hyperstress finds its field--theoretical
image in the densities of hypermomentum, i.e. in the
\begin{equation}
spin\>current\oplus dilation\>current\oplus shear\> current \>.
\end{equation}
And these currents ought to couple to the corresponding
post--Riemannian structures, a hypothesis which brought the
metric--affine gravity theory under way in the first place [He9,
He10].

According to Sakharov [Sa2, Sa3], gravitation represents a ``metrical
elasticity'' of space which is brought about by quantum fluctuations
of the vacuum. Here we pursue this analogy with continuum mechanics
much further and introduce additional nonmetric and torsional degrees
of freedom into spacetime, but, we believe, it is done in the same
spirit.

\section{Hadrons as extended structures --- effective `strong gravity'}

With the discovery of a spatial spread for the hadrons --- first in
experiments measuring the electromagnetic form factors, then in the
identification of the baryons with an SU(3) octet (rather than with
the fundamental representation of the group, as in the Sakata model)
and the conception of quarks as constituents --- it became important
to describe the dynamics and kinematics of quantum extended structures
(extendons).  The 1965 work dealt with three--dimensional vibrating
and rotating "lumps" [Do5]. Then came dual models [Ve1] and their
reinterpretation as a quantum string [Na1, Su3], a one--dimensional
extendon. It was later understood as an ``effective" description of
QCD flux tubes, extending between point--quarks [Ni4].

      Extendons can be deformed, and thus represent affine geometries
in themselves. Hadron excitations show up as Regge trajectories, and
the massive states fit $\overline{SL}(3,R)$ representations --- as
would indeed be expected from the pulsations of a (consider it as an
approximation) fixed--volume $3$--extendon [Do5]. The quantum
$d$--extendon involves covariance of a $d+1$ manifold (e.g. the
world--sheet for the string), the extendon's time evolution.  This
resembles gravity, involves gauging geometrical groups and
often reproduces the same equations that were derived in the pursuit
of quantum gravity.

      It is thus not surprising that ``effective strong--gravity"
theories, in which the Planck length $\ell$ is replaced by the Compton
wavelength of the proton, were derived in the same context. One such
example [He23] with a confining P\"oschl--Teller type potential [Mi6]
in the effective radial Schr\"odinger equation arose in the Poincar\'e
gauge theory and its generalization to $SL(6,C)$ flavor models of
Salam et al.[Sa5, Sa5a, Mi3a]. By including the $SU(3)$ color group of
QCD one ends up with the $SL(6,C)^{f}\otimes SL(6,C)^{c}$ model of
color geometrodynamics [Mi4, Mi6]. Another such treatment has used
affine manifolds [Ne14, Ne22].  In these models, ``low energy" means
"hadron energies", i.e.  1 -- 100 GeV.  The slope of the trajectories
is of the order of $1$ GeV, as against the $10^{19}$ GeV of the
theories we mentioned in Sec.1.1 above.

In a recent version of this approach, ``chromogravity'' [Ne22a, Ne22b,
Si8, Si9] is derived from QCD itself, as an ``effective'' theory. A
gravitation--like component is identified in the infrared limit of
QCD, its contribution providing for color confinement, for the
systematics of the excitations in the hadron spectrum (Regge
sequences) and for the forces of longer--range responsible for the
nuclear excitation spectrum. This QCD-generated graviton--like
component is the analog of van der Waals forces in molecules, where a
$J=2$ combination of two photons is exchanged between atoms; in QCD, a
$J=2$--mediated zero-color component, plus all higher spin zero-color
combinations of QCD gluons, make up this pseudo-gravitational
component.  The emergence of a $J=2$ contribution from a $J=1$ force
(QCD) in higher orders is similar to the generation of the $J=2$
gravitational contribution in string theory, from closed strings --
i.e. from the contraction of two open strings -- an open string
corresponding in the massless sector to a $J=1$--mediated force.

\section{The early universe (cosmogony)}

Already in the seventies, various theorems implied that, with a
cosmology based on Riemannian geometry, the universe was forced either
to have come out of a singularity --- or, inevitably, to fall into one
in the future.  The simplest way of avoiding such a result is to
assume that in the distant past --- or the distant future --- the
geometry is not Riemannian.

In the late seventies and in the eighties, the same conclusion emerged
from the new studies of the early universe connected with gauge
unification theories\footnote{Although the initials GUT were
originally taken to mean Grand Unified Theories, it was later agreed
(1979 HEPAP Conference) to read them as Gauge Unification Theories, in
order to leave room, as might be needed, for `grander' theories some
day.} (GUT) [Or0a] and their supersymmetric extensions -- later
replaced by unification and superunification as derived from the
quantum superstring.  In these theories the early universe has
additional dimensions (and superdimensions). It is assumed that these
extra dimensions spontaneously compactify, leaving internal symmetries
as residual effects in the final four-dimensional spacetime, cf.\
[We2b]. The symmetries that we have identified phenomenologically
include those of the $SU(3)\times SU(2) \times U(1)$ group of the
standard model embedded within higher rank groups such as $E(6)$ or
$SU(5)$.  All of this implies geometries ranging from K\"ahler and
Calabi-Yau to affine manifolds.

The eighties also ushered in {\it inflationary cosmology} [Gu1a, Li0,
Al2], a new conception of the very early universe, now from the point
of view of cosmology itself, rather than particle physics (though it
does affect it too). In the more advanced ``extended'' models [La1a]
one finds it necessary to abandon the Riemannian constraints [St2a],
at the very least replacing Einstein's geometry by Weyl's. We deal
with this situation in an example in Chap.\ 6.

\section{Organization of the paper and notation}

%
%
%
In Chap.2 we take a tour d'horizon around quantum gravity. We mention
the main open questions and unsolved problems.

In Chap.3 we show how, by starting with the affine group $A(n,R)$ and
its Yang--Mills type gauging, we eventually arrive at a metric--affine
geometry of spacetime, the structures and properties of which we
explicate in the rest of this chapter. In particular, the potentials
emerging from the affine connection are the coframe and the linear
connection. The latter is decomposed into Riemannian and
post--Riemannian pieces, and the interrelations of the Chern--Simons
terms to the Bianchi identities are exhibited. The rules of exterior
calculus we defer to Appendix A and the irreducible decompositions of
nonmetricity, torsion, curvature, and of the Bianchi identities to
Appendix B. All this is more or less traditional wisdom. However, we
stress the post--Riemannian structures in a coherent geometrical
framework, such as nonmetricity, the Weyl one--form, and the
volume--preserving piece of the connection.

In Chap.4 the question is answered of how one can present especially
{\it fermionic} matter in such a metric--affine spacetime. The results
of this chapter are fairly new and have been found during the last 15
years or so by one of us (YN) and his collaborators. World spinors are
defined and their conformal properties studied. Technical details of
the unitary irreducible representations of the $\overline{SA}(4,R)$
and the corresponding subgroups are collected in Appendix C.

Having now a spacetime arena available and matter fields `moving'
therein, we can build up a Lagrangian of this gravitationally
interacting matter system and an action function as well. This is done
in Chap.5 in the conventional way. We postulate affine gauge
invariance and switch on the Lagrange--Noether machinery. Besides the
conventional canonical energy--momentum current, we define,
generalizing the spin current, a hypermomentum current that is coupled
to the linear connection, i.e.\ to the new gravitational potential of
spacetime.

The Noether identities (5.2.10), (5.2.16) and the general form of the
gravitational field equations (5.5.2), (5.5.3), (5.5.4) are derived.
We discuss the Belinfante--Rosenfeld symmetrization of the
energy--momentum current and study different limiting cases of the
gravitational field equations by means of the Lagrange multiplier
technique. Finally Astekhar type complex variables are generated by
means of a metric--affine Chern--Simons term in the gauge Lagrangian.
Whereas most of the material of this chapter appeared before, we claim
some originality as to the {\it completeness} and the {\it rigor} of
our presentation.

Up to including Chap.5, no gravitational gauge Lagrangian is specified
explicitly. Thus we provided a `kinematical' framework for
metric--affine gauge gravity which has to be filled with physical
life. This is done in Chap.6. Conformally invariant gravitational
gauge Lagrangians, including dilaton fields, are studied and compared
to alternative approaches in the literature. Various schemes of
symmetry reduction from the linear to the Lorentz group are given
explicitly. The exact procedure is one of the main open problems.  We
believe, however, that the solution of exactly this problem is
indispensable for future progress in gravity. Moreover, we discuss
generalizations of recent inflationary models in our post--Riemannian
framework.

The list of literature is organized so as to be of maximal
usefulness to the reader. We tried to refer to all material relevant
to our task. Should we have overlooked some articles, we would like to
ask the authors to let us know, possibly by email to
hehl@thp.uni-koeln.de. We may want to supply this additional
information in an erratum.

In the body of the paper, the gravitational models used will be
abbreviated as follows:

\noindent {\bf GR} = general relativity theory, also called Einstein
gravity (Riemannian spacetime $V_4$) [Ei0].

\noindent {\bf GR$_\parallel$} = teleparallel (version of general
relativity) theory (Weitzenb\"ock spacetime $W_4$: Riemann--Cartan
spacetime with vanishing (Cartan--)curvature and non--vanishing
torsion), see [Ni6b, Sc11].

\noindent {\bf EC theory} = Einstein--Cartan(--Sciama--Kibble
theory of) gravity (Riemann--Cartan spacetime $U_4$: Metric and
metric--compatible connection), see [Tr2].

\noindent {\bf PG} (theory) = Poincar\'e (gauge theory of) gravity
(Riemann--Cartan spacetime $U_4$), see [He4].

\noindent {\bf MAG} = Metric-affine (gauge theory of) gravity
(metric--affine spacetime $(L_4,g)$: Independent $GL(4,R)$--connection
and independent metric), see [He19].

\noindent We denote the covering of a certain group by an overline. We have,
for instance, $SL(2,C)=\overline{SO}(1,3)$. Sometimes we dispense with
the overline for convenience provided it is clear from the context anyways.

\section{Acknowledgments}

This paper was only made possible through substantial support of the
German-Israeli Foundation for Scientific Research \& Development
(GIF), Jerusalem \& Munich.

Different people helped us at various stages of the writing up of the
paper. We are most grateful to all of them. Yuri Obukhov
(Moscow/Cologne) read very carefully a preliminary version of our
article and came up with numerous suggestions. Djordje \v Sija\v cki
(Belgrade) as well as Tom Laffey (Dublin) and J\"urgen Lemke
(Cologne/Austin) were of great help in group--theoretical questions.
J\"org Hennig (Clausthal) advised us on bundles, Norbert Straumann
(Z\"urich) on densities, Ralf Hecht (Cologne/Chung-Li) on energy
complexes, Horst Konzen (Cologne) checked some of the algebra,
Romulado Tresguerres (Madrid/Cologne) contributed to our understanding
of conformal transformations, and Franz Schunck (Cologne) developed
some cosmological models. C.Y. Lee (Seoul) shared with us his
quantization experience and J. Godfrey (then Tel-Aviv) his kowledge of
projective geometry.

And last but not least, Dietrich Stauffer (Cologne/Antigonish)
promoted this project generously by his leave of absence from Cologne,
paid by the Canada Council. One of us (Y.N.) was Fall 1993 Joint Royal
Society/Israel Academy of Sciences and Humanities Research Professor
and would like to thank Prof.\ D.\ Lynden-Bell and the University of
Cambridge Institute of Astronomy for hospitality during the final
stages of this work.

\end{document}